\documentclass{jfm}
\usepackage{graphicx}
\usepackage{epstopdf, epsfig}
\usepackage{amsmath}
\usepackage{color}
\usepackage{booktabs}

\renewcommand{\d}{\mathrm{d}}

\newcommand{\uu}{\boldsymbol{u}}

\newcommand{\ff}{\boldsymbol{f}}

\newcommand{\ii}{\textrm{i}}

\title{Streamwise-constant large-scale structures in Couette and Poiseuille flows}
\shorttitle{Streamwise-constant large-scale structures in channel flows}

\shortauthor{S. J. Illingworth}
\author{
  Simon J. Illingworth
  \corresp{\email{sillingworth@unimelb.edu.au}}
  }

\affiliation{
  Mechanical Engineering,
  University of Melbourne,
  VIC 3010, Australia
}

\begin{document}

\maketitle

\begin{abstract}
The linear amplification mechanisms leading to streamwise-constant large-scale structures in laminar and turbulent channel flows are considered.
   A key feature of the analysis is that the Orr--Sommerfeld and Squire operators are each considered separately.
  Physically this corresponds to considering two separate processes: (i) the response of wall-normal velocity fluctuations to external forcing;
  and (ii) the response of streamwise velocity fluctuations to wall-normal velocity fluctuations. 
  In this way we exploit the fact that, for streamwise-constant fluctuations, the dynamics governing the wall-normal velocity are independent of the mean velocity profile (and therefore the mean shear).
  The analysis is performed for both plane Couette flow and plane Poiseuille flow; and for each we consider linear amplification mechanisms about both the laminar and turbulent mean velocity profiles.
  The analysis reveals two things.
  First, that the most amplified structures (with a spanwise spacing of approximately $4h$, where $h$ is the channel half height) are to a great extent encoded in the Orr-Sommerfeld operator alone, thus helping to explain their prevalence.
  Second---and consistent with numerical and experimental observations---that Couette flow is significantly more efficient than Poiseuille flow in leveraging wall-normal velocity fluctuations to produce large-scale streamwise streaks.
\end{abstract}


\section{Introduction}
Streamwise-elongated large-scale structures are prevalent in all of the canonical wall-bounded shear flows: boundary layers \citep{adrian2000voo,balakumar2007lav,hutchins2007evl}; pipes \citep{kim1999vls,guala2006lsa,monty2007lsf,monty2009ctp}; Poiseuille flow \citep{balakumar2007lav,monty2007lsf,monty2009ctp}; and Couette flow.
In this context Couette flow is peculiar in that streamwise-elongated channel-wide structures are especially pronounced, evidence of which has been observed across a broad range of Reynolds numbers both in simulations \citep{lee1991sts,bech1994vls,komminaho1996vls,papavassiliou1997ils,tsukahara2006dns,avsarkisov2014tpc,pirozzoli2014tsc,lee2018esm} and in experiments \citep{tillmark1994,bech1994vls,tillmark1998lss,kitoh2005esm,kitoh2008esl}.

Meanwhile linear analyses of the Navier--Stokes equations also uncover an important place for streamwise-elongated structures in shear flows \citep{schmid2001sts}.
Such linear analyses reveal that, for channel flows (in which we include both Poiseuille flow and Couette flow), the structures that are most excitable are streamwise-constant with a spanwise spacing of approximately $4h$, corresponding to a spanwise wavenumber of approximately $\pi/2h$ (where $h$ is the channel half height).
This has been observed not only for laminar Couette flow and laminar Poiseuille flow \citep{gustavsson1991egt,butler1992tdo,farrell1993sfl,trefethen1993hsw,jovanovic2005cea}, but also more recently for their turbulent counterparts for which the linear analyses are performed about the turbulent mean flow \citep{alamo2006lea,pujals2009not,hwang2010acs,hwang2010lnn}.
A key ingredient in the development of these streamwise-constant structures is the lift-up effect, the driving mechanism for which is the mean wall-normal shear \citep{ellingsen1975slf,landahl1980note,kim2000lpw}.
We can therefore summarize as follows: mean shear is a key ingredient in linear amplification mechanisms; and yet the most amplified structures in channel flows are largely insensitive to the details of the shear.

This paper considers the linear amplification mechanisms leading to streamwise-constant large-scale structures in laminar and turbulent channel flows.
The importance of streamwise-constant structures in channel flows has motivated a number of previous investigations using both linear \citep{bamieh2001eac,jovanovic2005cea} and nonlinear \citep{gayme2010scm,gayme2011anm} modelling approaches.
A key feature of the present analysis is that the Orr--Sommerfeld and Squire operators are each considered separately.
Physically this corresponds to considering two separate processes: (i) the response of wall-normal velocity fluctuations to external forcing;
and (ii) the response of streamwise velocity fluctuations to wall-normal velocity fluctuations. 
In this way we exploit the fact that, for streamwise-constant fluctuations, the dynamics governing the wall-normal velocity are independent of the mean velocity profile (and therefore the mean shear).
This point of view, in which the forcing of streamwise velocity by wall-normal velocity is made explicit, is in the spirit of \citet{gustavsson1991egt}.
The analysis is performed for both plane Couette flow and plane Poiseuille flow; and for each we consider linear amplification mechanisms about both the laminar and turbulent mean velocity profiles.
By doing so we will uncover two things.
First, that the most amplified structures---with a spanwise spacing of approximately $4h$ irrespective of the details of the mean flow---are to a great extent encoded in the Orr-Sommerfeld operator alone, thus helping to explain the prevalence of such structures.
Second---and consistent with numerical and experimental observations---that Couette flow is significantly more efficient than Poiseuille flow in leveraging wall-normal velocity fluctuations to produce large-scale streamwise streaks.

\section{Linear model} \label{sec:model}
We consider laminar or turbulent flow in a channel for which the streamwise, spanwise and wall-normal directions are denoted by $x$, $y$ and $z$; and the corresponding velocity components by $u$, $v$ and $w$.
The Reynolds number $R = u_\textrm{o} h / \nu$ is based on the channel half-height $h$, kinematic viscosity $\nu$, and some characteristic velocity $u_\textrm{o}$.
For laminar flow this characteristic velocity is the maximum velocity across the channel height; for the turbulent flow it is the friction velocity, $u_\tau = \sqrt{\tau_w/\rho}$, where $\tau_w$ is the wall shear stress and $\rho$ is the density.
Following non-dimensionalization the channel half-height is unity so that $z \in [-1,+1]$.

\subsection{Laminar velocity profiles} \label{sec:lamProf}
The non-dimensional velocity profile for laminar Couette flow is $U(z) = z$; for laminar Poiseuille flow it is $U(z) = 1-z^2$.
Linearizing the incompressible Navier--Stokes equations about one of these laminar base flows gives
\begin{subequations} \label{eq:lns}
  \begin{gather}
    \frac{\p \uu}{\p t} + U \frac{\p \uu}{\p x}
    + (wU',0,0)
    =
    -\nabla p + R^{-1} \Delta \uu + \ff \label{eq:mom} \\
    \nabla \cdot \uu = 0, \label{eq:cont}
  \end{gather}
\end{subequations}
where $\uu = [u \ v \ w]^T$ and $'$ represents differentiation in the wall-normal direction.
Note the inclusion of a forcing term, $\ff = [f_x \ f_y \ f_z]^T$, in the momentum equation \eqref{eq:mom}, which we treat as an external input to the flow.

\subsection{Turbulent velocity profiles}
Linear models have been used for fully developed turbulent flows in a number of previous studies.
In some the linear operator is obtained from a Reynolds decomposition of the velocity field, giving rise to a linear operator in which the viscosity is equal to the kinematic viscosity \citep{mckeon2010clf,sharma2013ocs}.
In others the linear operator is obtained by first performing a triple decomposition of the velocity field and then providing a closure for the terms quadratic in the incoherent fluctuations using a simple eddy viscosity model \citep{reynolds1972mow,alamo2006lea,pujals2009not,hwang2010acs,hwang2010lnn,illingworth2018els,vadarevu2019csl}.
In this second case the effective viscosity, which varies across the flow, is given by the sum of the eddy and kinematic viscosities.

In this work we include only the kinematic viscosity in the linear model.
Doing so gives rise to an Orr--Sommerfeld operator whose dynamics are independent of the mean velocity profile (to be made clear in \S\,\ref{sec:streamwise}) and thus simpler and more generic than its eddy-viscosity equivalent.
This choice is therefore convenient but it is also suitable for two reasons.
First, a key emergent feature of both linear models is the critical layer mechanism in which a significant response can occur when the phase velocity, $c = \omega/k_x$, is equal to the local mean velocity, $c = U(z)$ \citep{maslowe1986critical}.
The two linear models show different critical layer behaviours owing to their different effective viscosity profiles.
But since the focus of this work is on streamwise-constant fluctuations for which the streamwise wavenumber is zero, the critical layer mechanism does not occur and this difference between the two models does not exist.
Second, it is necessary to know the variation of the eddy viscosity across the flow, but a reasonable approximation is to assume that it is constant.
As noted by \citet[\S\,6.7 p. 127]{townsend1956sts}, a reasonable way to determine this constant would be to compare the measured mean velocity profile with that given by assuming a suitable constant eddy viscosity; and the approximation would only be in error near the walls.
It is therefore reasonable to approximate any eddy viscosity with an equivalent constant (and larger) viscosity which would manifest itself as a reduction in the effective Reynolds number $R$.

The linear model about the turbulent mean flow has the same form as that for laminar flow \eqref{eq:lns} but the reasoning used to form it and the definition and interpretation of some its terms are different.
The linear model is obtained by performing a Reynolds decomposition of the velocity field; substituting this into the incompressible Navier--Stokes equations; and subtracting the mean equations.
This gives rise to \eqref{eq:lns} as before but with two important differences.
First, $\boldsymbol{U}$ now represents the turbulent mean velocity profile and $\uu$ represents turbulent fluctuations about the mean.
Second, any nonlinear terms are absorbed into the forcing term $\ff$ so that $\ff = - (\uu \cdot \nabla) \uu + \overline{(\uu \cdot \nabla) \uu}$ \citep{mckeon2010clf}.
Thus the formation of the linear operator \eqref{eq:lns} about the turbulent mean flow does not imply that nonlinear terms are neglected.
Rather we take the point of view that the linear operator \eqref{eq:lns} is constantly forced by the remaining nonlinear terms.
Nonlinear effects thus manifest themselves in two ways: through their role in setting the mean velocity profile $\boldsymbol{U}$; and through their forcing of the linear operator \eqref{eq:lns}.
(The forcing could have additional contributions from any externally applied forcing as in \S\,\ref{sec:lamProf}, but this would not modify the analysis.)

\begin{figure} \centering
  \includegraphics[width=0.8\textwidth]{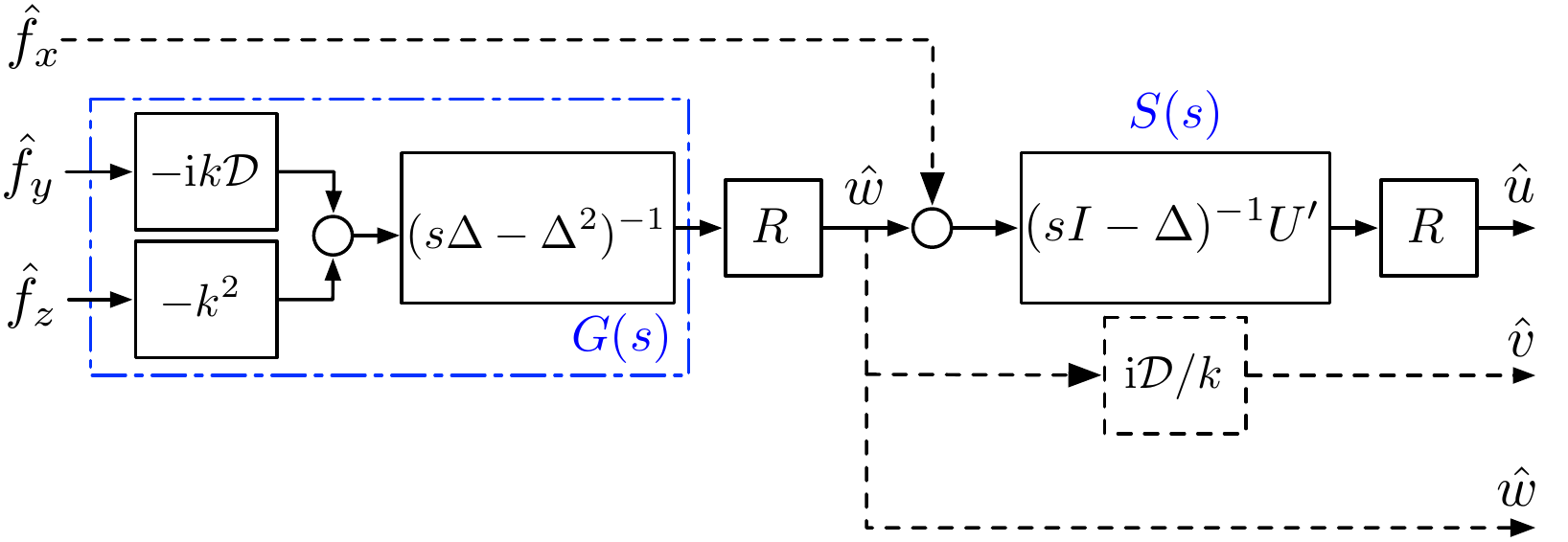}
  \caption{\label{fig:kx0}
    Linear amplification mechanisms for streamwise-constant fluctuations at spanwise wavenumber $k$.
    Dashed lines denote forcing components and velocity components that are omitted in the analysis.
    (These processes scale with $R$, while those retained scale with $R^2$---see \S\,\ref{sec:streamwise}.)
    The transfer functions $G(s)$ from \eqref{eq:G} and $S(s)$ from \eqref{eq:S} are also indicated.
  }
\end{figure}

\subsection{Streamwise-constant model} \label{sec:streamwise}
The streamwise-constant model used throughout this work is obtained by taking Fourier transforms of \eqref{eq:lns} in the homogeneous directions ($x$,\,$y$); transforming into Orr--Sommerfeld Squire form; and setting the streamwise wavenumber to zero:
\begin{subequations} \label{eq:wu}
\begin{gather} 
  \frac{\p \Delta \hat{w}}{\p t} = R^{-1} \Delta^2 \hat{w} - \ii k \mathcal{D} \hat{f}_y - k^2 \hat{f}_z \label{eq:w} \\
  \frac{\p \hat{u}}{\p t} = R^{-1} \Delta \hat{u} - U' \hat{w} + \hat{f}_x, \label{eq:u}
\end{gather}
\end{subequations}
with boundary conditions $\hat{w}(t) = \p_z\hat{w}(t) = \hat{u}(t) = 0$ at the two walls.
Here $k$ is the spanwise wavenumber, $\mathcal{D}$ represents differentiation in the wall-normal direction, and $\Delta = \mathcal{D}^2-k^2$ is the two-dimensional Laplacian.
Note that we choose an evolution equation for wall-normal velocity fluctuations $\hat{w}$ in \eqref{eq:w} instead of $\hat{v}$ (or a suitable streamfunction) because $\hat{w}$ forces $\hat{u}$ directly in \eqref{eq:u}.
$\hat{v}$ can be obtained from $\hat{w}$ by using the continuity equation \eqref{eq:cont} with streamwise gradients set to zero (see also figure \ref{fig:kx0}):
\begin{equation}
  \ii k \hat{v} + \mathcal{D} \hat{w} = 0.
\end{equation}
Note also that, rather than using the (standard) wall-normal vorticity as the output of the Squire operator in \eqref{eq:u}, we use streamwise velocity fluctuations, $\hat{u}$, since this is the quantity of interest.

Taking Laplace transforms of \eqref{eq:wu}, rescaling the Laplace variable $s$ by the Reynolds number $R$ \citep{gustavsson1991egt,jovanovic2005cea} and rearranging gives rise to two transfer functions, $G(s)$ and $S_\textrm{o}(s)$, as follows:
\begin{subequations} \label{eq:GS0}
  \begin{align}
    \hat{w}(s) &= R \underbrace{(s\Delta - \Delta^2)^{-1}
    \begin{bmatrix} -\ii k \mathcal{D} & -k^2 \end{bmatrix}}_{G(s)}
    \begin{bmatrix} \hat{f}_y(s) \\ \hat{f}_z(s) \end{bmatrix} \label{eq:G}
    \\
    \hat{u}(s) &= R \underbrace{(sI - \Delta)^{-1}
    \begin{bmatrix} I & -U' \end{bmatrix}}_{S_\textrm{o}(s)}
    \begin{bmatrix} \hat{f}_x(s) \\ \hat{w}(s) \end{bmatrix}. \label{eq:S0}
  \end{align}
  (Rescaling the Laplace variable in this way is equivalent to rescaling time by $R^{-1}$.)
  Notice that the Reynolds number $R$ has been isolated to render each transfer function independent of it.
  The transfer function $G(s)$ comes from the Orr-Sommerfeld operator.
  It represents the dynamics from spanwise and wall-normal forcing, $[\hat{f}_y \ \hat{f}_z]^T$, to wall-normal velocity fluctuations, $\hat{w}$.
  Importantly, $G(s)$ is independent of the mean velocity profile.
  The transfer function $S_\textrm{o}(s)$ comes from the Squire operator.
  It represents the dynamics from streamwise forcing and wall-normal velocity fluctuations, $[\hat{f}_x \ \hat{w}]^T$, to streamwise velocity fluctuations, $\hat{u}$.
  Equations \eqref{eq:GS0} are represented in a block diagram in figure \ref{fig:kx0}.
  
  From (\ref{eq:G},~\ref{eq:S0}) and figure \ref{fig:kx0} we see that the overall transfer function from $[\hat{f}_y \ \hat{f}_z]^T$ to $\hat{u}$ (which involves both $G(s)$ and $S_\textrm{o}(s)$) scales with $R^2$, while the transfer function from $\hat{f}_x$ to $\hat{u}$ (which involves only $S_\textrm{o}(s)$) scales with $R$.
  For sufficiently large $R$ we can therefore neglect the influence of $\hat{f}_x$ and it is convenient to define a second transfer function, $S(s)$, for which the influence of $\hat{f}_x$ is neglected:
  \begin{equation} \label{eq:S}
    \hat{u}(s) = R \underbrace{[-(sI - \Delta)^{-1}U']}_{S(s)} \hat{w}(s).
  \end{equation}
\end{subequations}
The transfer function $S_\textrm{o}(s)$ in \eqref{eq:S0} therefore includes the influence of $\hat{f}_x$, while the transfer function $S(s)$ in \eqref{eq:S} neglects it.

Finally we define an overall transfer function $F(s)$:
\begin{equation} \label{eq:F}
  \begin{aligned}
    F(s)
    &= S(s) G(s) \\
    &= (sI - \Delta)^{-1} U' (s\Delta - \Delta^2)^{-1}
    \begin{bmatrix} \ii k \mathcal{D} & k^2 \end{bmatrix}.
  \end{aligned}
\end{equation}
Implicit in \eqref{eq:F} is that the contribution from streamwise forcing $\hat{f}_x$ can be ignored.
This assumption is sound provided that the Reynolds number $R$ is sufficiently large as discussed above.
Then $F(s)$ represents the overall dynamics from $[\hat{f}_y \ \hat{f}_z]^T$ to $\hat{u}$ as follows:
\begin{equation} \label{eq:f2u}
  \hat{u}(s) = R^2 F(s)
  \begin{bmatrix} \hat{f}_y(s) \\ \hat{f}_z(s) \end{bmatrix}.
\end{equation}
The overall dynamics \eqref{eq:f2u}, together with definitions of $G(s)$ and $S(s)$, are shown in a block diagram in figure \ref{fig:kx0}.
  
\subsection{Transfer function norms} \label{sec:norms}
With the relevant transfer functions defined, we now introduce two norms to evaluate their size: the infinity norm $\|\cdot\|_\infty$ and the two norm $\|\cdot\|_2$.

We will make extensive use of the infinity norm, defined for a transfer function $T$ as
\begin{align} \label{eq:infnorm}
  \|T\|_\infty = \max_\omega \, \sigma_1(\ii \omega).
\end{align}
$\sigma_i(\ii \omega)$ are the singular values of $T(\ii\omega)$ at frequency $\omega$ and represent a generalization of gain for systems with many inputs and many outputs.
The singular values are ordered such that $\sigma_1 \ge \sigma_2 \ge \ldots \ge \sigma_n$.
Therefore $\sigma_1$ represents the maximum singular value at frequency $\omega$; and the infinity norm \eqref{eq:infnorm} represents the worst-case gain over all possible forcing frequencies and forcing directions.
An important property of the infinity norm---and a key reason for using it in this work---is its submultiplicative property:
\begin{equation} \label{eq:submult}
  \|T_1 T_2\|_\infty \le \|T_1\|_\infty \|T_2\|_\infty
\end{equation}
for any two transfer functions $T_1$ and $T_2$.
This property will be useful in the following sections to characterize the efficiency of the forcing of streamwise velocity fluctuations (the Squire operator $S$) by wall-normal velocity fluctuations (the Orr--Sommerfeld operator $G$).

Another commonly used norm for transfer functions is the two norm:
\begin{align} \label{eq:2norm}
  \|T\|_2^2
  = \frac{1}{2\pi} \int_{-\infty}^\infty \textrm{Trace} [ T^*(\ii\omega) T(\ii\omega) ] \, \d \omega
  = \frac{1}{2\pi} \int_{-\infty}^\infty \sum_i \sigma_i^2(\ii\omega) \, \d \omega.
\end{align}
The two norm \eqref{eq:2norm} represents an average gain over all frequencies and forcing directions.
In contrast to the infinity norm, the two norm does not satisfy the submultiplicative property \eqref{eq:submult}, and is therefore less suitable for our purposes (to be made clear in \S\,\ref{sec:eff}).
Nevertheless it will be used briefly in \S\,\ref{sec:laminar} to check the sensitivity of some key results to the choice of norm.

\subsection{Numerical discretization}
Equations \eqref{eq:G} and \eqref{eq:S} are discretized in the wall-normal direction ($z$) using Chebyshev collocation.
101 Chebyshev points are used, which is sufficient for the large scales of interest.
Convergence has been checked by doubling the number of Chebyshev points and ensuring that the results do not change.
The norms \eqref{eq:infnorm} and \eqref{eq:2norm} are defined such that they each correspond to grid-independent energy norms.
This is achieved using Clenshaw--Curtis quadrature \citep{trefethen2008gauss}.

\section{Laminar velocity profiles} \label{sec:laminar}
We start by evaluating the infinity norm \eqref{eq:infnorm} for laminar Couette flow and laminar Poiseuille flow as a function of the spanwise wavenumber $k$.
We do this for the Orr--Sommerfeld operator $G$, the Squire operator $S$, and for the overall system $F=SG$.

\subsection{Norms of $G$, $S$ and $F$ with spanwise wavenumber} \label{sec:normk}
\begin{figure} \centering
  \includegraphics[width=1\textwidth]{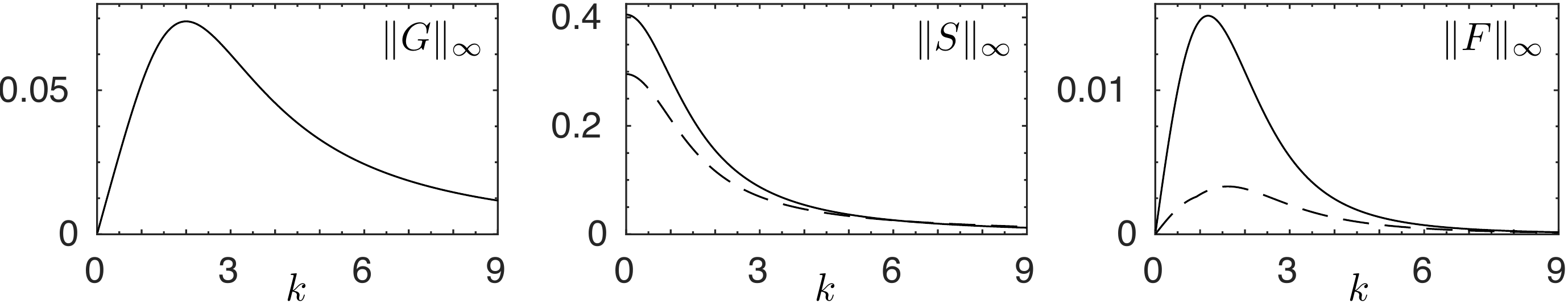}
  \caption{\label{fig:infnorm}
    The infinity norm $\|\cdot\|_\infty$ for $G$ (left); $S$ (middle); and $F=SG$ (right).
    Results for $S$ and $F$ are shown for Couette flow (---) and for Poiseuille flow ($--$).
  }
\end{figure}

Figure \ref{fig:infnorm} plots the infinity norm \eqref{eq:infnorm} as a function of spanwise wavenumber for $G$, $S$ and $F$.
The infinity norm of the Orr--Sommerfeld operator $G$ is plotted in figure \ref{fig:infnorm}\,(\textit{a}).
Recall that, for streamwise-constant fluctuations, the dynamics of $G$ are independent of the mean velocity profile and are therefore identical for Poiseuille flow and Couette flow.
We observe that $\|G\|_\infty$ attains its maximum at a spanwise wavenumber of $k=2.00$ ($\lambda = 3.14$).
The infinity norm of the Squire operator $S$ is plotted in figure \ref{fig:infnorm}\,(\textit{b}).
Notice that there are now two curves---one for Couette flow and one for Poiseuille flow---because $S$ is a function of the mean velocity profile.
For both flows $\|S\|_\infty$ attains its maximum at $k=0$ and decreases monotonically with increasing $k$.
Finally the infinity norm of $F=SG$, which represents the overall dynamics, is plotted in figure \ref{fig:infnorm}\,(\textit{c}).
For both flows $\|F\|_\infty$ attains its maximum near $k \approx \pi/2$ ($\lambda \approx 4$).
Precisely, $\|F\|_\infty$ attains its maximum at $k=1.18$ ($\lambda=5.3$) for Couette flow and at $k=1.62$ ($\lambda = 3.9$) for Poiseuille flow.

What if we use a different norm?
Table \ref{tab:kmax} summarizes the spanwise wavenumbers at which the peak response, as measured by the infinity norm, is achieved; and also repeats this using the two norm \eqref{eq:2norm}.
(The variation of the two norm with $k$ is plotted in appendix \ref{app:2norm}.)
From table \ref{tab:kmax} and appendix \ref{app:2norm} we see that the responses of the operators $G$, $S$ and $F$, as measured by the 2 norm, show similar behaviour to that seen using the infinity norm.
In particular we still observe i) that $G$ and $F$ each attain their maximum response at a spanwise wavenumber $k \approx 2$; and ii) that $S$ attains its maximum response at $k=0$.
Thus we see that pertinent results observed when using the infinity norm are also observed when using the two norm, and therefore that these features are not an artefact of our particular choice of norm.

\begin{table}
  \begin{center}
    \def~{\hphantom{0}}
    \begin{tabular}{lccccc}
      &  $G$  &  $S_\textrm{coue}$  &  $S_\textrm{pois}$  &   $F_\textrm{coue}$  &  $F_\textrm{pois}$ \\
      \midrule
      $k_\textrm{max}$ for $\|\cdot\|_\infty$ \quad \phantom{} & 2.00 & 0 & 0 & 1.18 & 1.62 \\
      $k_\textrm{max}$ for $\|\cdot\|_2$                        & 2.90 & 0 & 0 & 1.40 & 1.79 \\
    \end{tabular}
    \caption{Spanwise wavenumber $k_\textrm{max}$ at which the largest response occurs for the infinity norm, $\|\cdot\|_\infty$, and the two norm, $\|\cdot\|_2$.}
    \label{tab:kmax}
  \end{center}
\end{table}

From figure \ref{fig:infnorm} it is interesting that, despite the norm $\|G\|_\infty$ remaining unchanged and the norm $\|S\|_\infty$ remaining similar across the two flows, the norm of the overall system, $\|F\|_\infty$ differs significantly.
In particular $\|F\|_\infty$ for Couette flow is significantly larger than $\|F\|_\infty$ for Poiseuille flow.
We now consider this in more detail by defining an efficiency of the forcing process.

\subsection{Forcing efficiency} \label{sec:eff}
We now make use of the submultiplicative property of the infinity norm (see \S\,\ref{sec:norms}):
\begin{equation} \label{eq:infnorm_ineq}
  \| S G\|_\infty \le \|S\|_\infty \|G\|_\infty.
\end{equation}
In words: The optimal response of the overall system, when quantified using the infinity norm, is at most as large as the product of the optimal responses of the two systems of which it is composed.
How closely equality in \eqref{eq:infnorm_ineq} is approached is determined by the nature of the interaction between $G$ and $S$.
With \eqref{eq:infnorm_ineq} in mind we now introduce a forcing efficiency, $\alpha$, defined such that
\begin{equation} \label{eq:infnorm_alpha}
  \| S G\|_\infty = \alpha \|S\|_\infty \|G\|_\infty,
\end{equation}
or 
\begin{equation} \label{eq:alpha}
  \alpha = \frac{\| S G\|_\infty}{\|S\|_\infty \|G\|_\infty}.
\end{equation}
Note that \eqref{eq:infnorm_ineq} and \eqref{eq:infnorm_alpha} together imply that $\alpha \le 1$.
A value of $\alpha=1$ implies that the forcing of streamwise velocity fluctuations by wall-normal velocity fluctuations is perfectly efficient.
A value of $\alpha \approx 0$ implies a low efficiency or that wall-normal velocity fluctuations ($G$) are not able in turn to easily excite streamwise velocity fluctuations ($S$).

\begin{figure} \centering
  \includegraphics[width=0.9\textwidth]{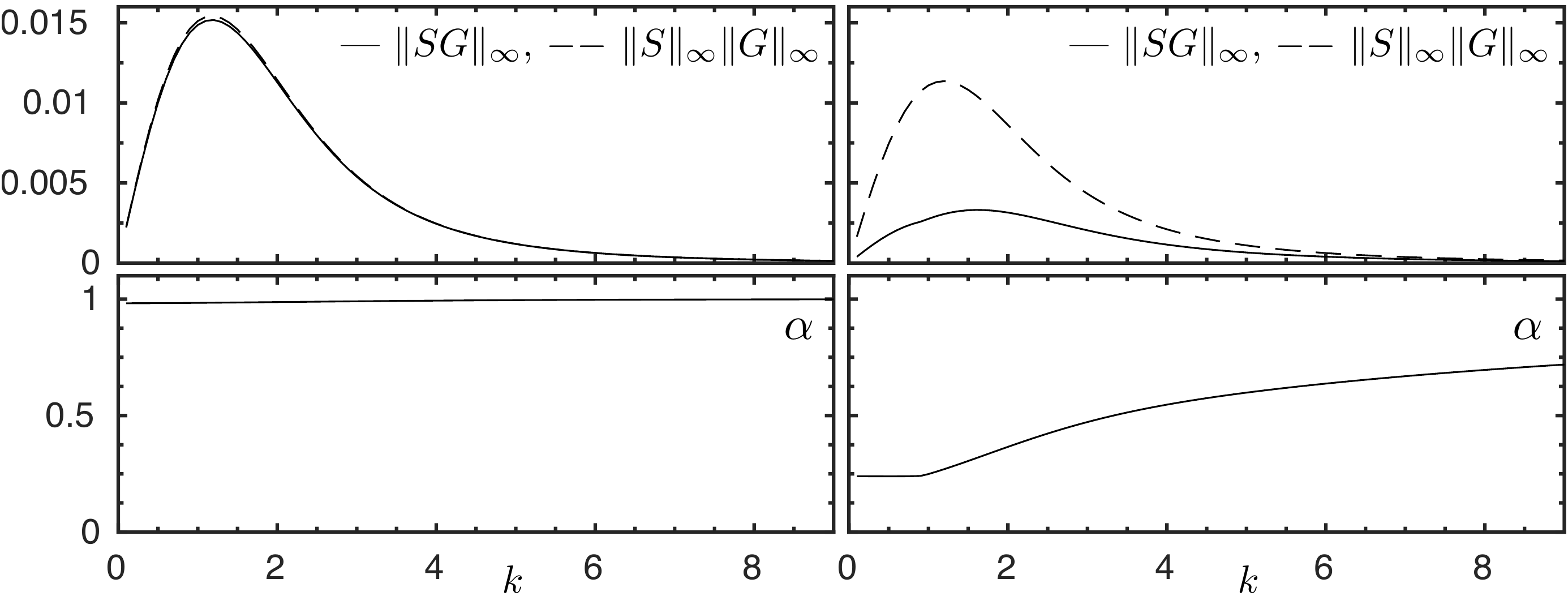}
  \caption{\label{fig:conversion}
    The infinity norm of the overall system, $\|SG\|_\infty$, and its upper bound $\|S\|_\infty\|G\|_\infty$; and the corresponding forcing efficiency $\alpha$ (see eq.~\eqref{eq:alpha}).
    Results are shown for Couette flow (left) and for Poiseuille flow (right).
  }
\end{figure}

Figure \ref{fig:conversion} shows the numerator and denominator of \eqref{eq:alpha}, together with the efficiency $\alpha$, for both Couette flow and Poiseuille flow.
From \eqref{eq:infnorm_ineq} we expect that the denominator, $\|S\|_\infty\|G\|_\infty$, will serve as an upper bound for the numerator, $\|SG\|_\infty$ and this is confirmed in panels (\textit{a},\textit{b}).
For Couette flow the upper bound is very nearly attained (panel (\textit{a})) corresponding to an efficiency $\alpha$ close to 1 (panel (\textit{b})).
For Poiseuille flow the efficiency is noticeably less than 1 at all spanwise wavenumbers considered and is in the region of $\alpha \approx 0.4$ for the spanwise wavenumbers at which the largest overall response is attained (i.e.~$k \approx \pi/2$).

\section{Turbulent velocity profiles} \label{sec:turbulent}
We now repeat the analysis of \S\,\ref{sec:laminar} using turbulent mean velocity profiles for both Couette flow and Poiseuille flow.
The focus of the results is similar to that of \S\,\ref{sec:laminar}, with the caveat that the mean velocity profile---and therefore the models $S$ and $F$---now vary with Reynolds number.
(Recall from \S\,\ref{sec:streamwise} that the Reynolds number has been eliminated from the operators $G$, $S$ and $F$; therefore the turbulent Reynolds number manifests itself only through the modification of the mean velocity profile.)
The turbulent mean velocity profiles are taken from existing turbulence databases \citep{hoyas2006scaling,hoyas2008reynolds,pirozzoli2014tsc,pirozzoli2017mct}.
The Reynolds numbers considered are summarized in table \ref{tab:re}.
For Couette flow we use data for friction Reynolds numbers $R_\tau$ between 171 and 986.
For Poiseuille flow we use data for friction Reynolds numbers between 107 and 2003.

\begin{table}
  \begin{center}
    \def~{\hphantom{0}}
    \begin{tabular}{rlrrrrrrr}
      \midrule
      Couette: & $R_\tau$    &      &  171 &  260 &   507 &       &   986 &       \\
      \midrule
      Poiseuille: & $R_\tau$ &  107 &  180 &  298 &   550 &   816 &   950 &  2003 \\
      \midrule
    \end{tabular}
    \caption{Turbulent Reynolds numbers considered.}
    \label{tab:re}
  \end{center}
\end{table}

\begin{figure} \centering
  \includegraphics[width=0.9\textwidth]{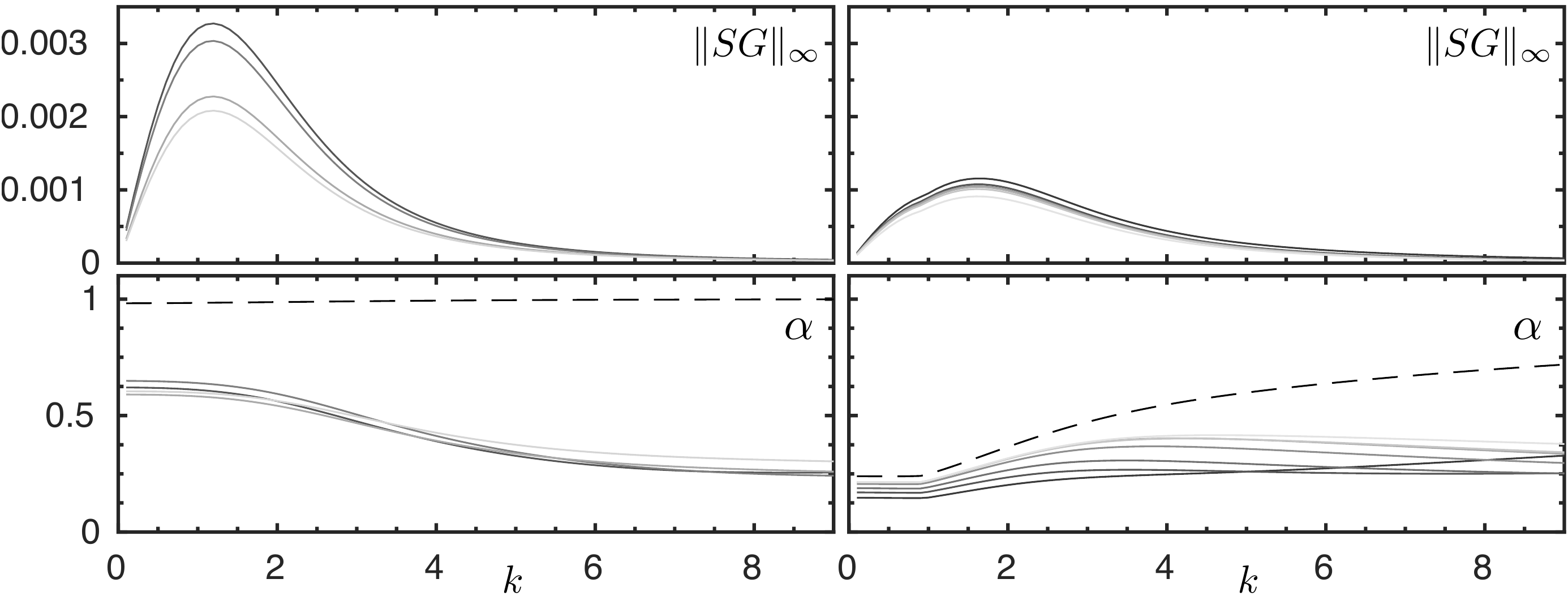}
  \caption{\label{fig:conversionRe}
    The infinity norm of the overall system, $\|SG\|_\infty$, as a function of the spanwise wavenumber $k$;
    and the efficiency, $\alpha$ as defined in \eqref{eq:alpha}.
    Results are shown for Couette flow (left) and for Poiseuille flow (right).
    Lighter lines correspond to larger friction Reynolds numbers $R_\tau$.
    For comparison the efficiencies $\alpha$ for laminar flow from figure \ref{fig:conversion} are also plotted ($--$).
  }
\end{figure}

In figure \ref{fig:conversionRe} we plot (for each Reynolds number) the infinity norm of the overall system, $\|F\|_\infty = \|SG\|_\infty$, as a function of the spanwise wavenumber $k$.
We also plot for each Reynolds number the efficiency, $\alpha$, as defined in \eqref{eq:alpha}.
For comparison we include the plots of $\alpha$ for laminar flow from figure \ref{fig:conversion} (dashed lines).
($\|S\|_\infty \|G\|_\infty$ is not plotted but can be inferred from knowledge of $\|SG\|_\infty$ and $\alpha$.)
We see that, for both flows and for all Reynolds numbers, a peak in $\|SG\|_\infty$ occurs for spanwise wavenumbers in the vicinity of $k = \pi/2$, as it did for laminar flow in figure \ref{fig:conversion}, and consistent with previous work \citep{pujals2009not,hwang2010acs,hwang2010lnn}.
This is perhaps not surprising: recall from figure \ref{fig:infnorm} that this peak is present in $G(s)$ governing the wall-normal velocity, which from \eqref{eq:G} is independent of the mean velocity profile and therefore has identical dynamics across all laminar and turbulent flows.
We also see in figure \ref{fig:conversionRe} that the efficiencies, $\alpha$, are lower than their laminar counterparts for all of the turbulent Reynolds numbers considered.

\begin{figure} \centering
  \includegraphics[width=0.95\textwidth]{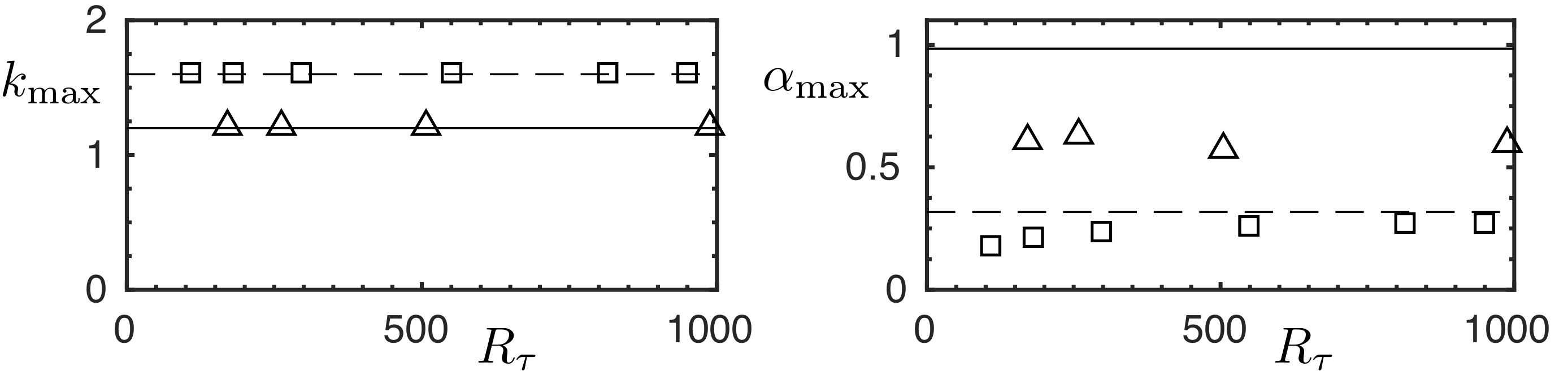}
  \caption{\label{fig:effRe}
    Summary of key results from figure \ref{fig:conversionRe} as a function of friction Reynolds number $R_\tau$:
    (left) spanwise wavenumber, $k_\textrm{max}$, at which $\|F\|_\infty$ attains its maximum; and (right) the efficiency $\alpha_\textrm{max}$ attained at this maximum.
    Results are shown for turbulent Couette flow ($\triangle$), turbulent Poiseuille flow ($\square$) and their laminar counterparts (Couette ---, Poiseuille $--$).
}
\end{figure}

Key results from figure \ref{fig:conversionRe} are summarized as a function of Reynolds number in figure \ref{fig:effRe}.
We plot in panel (\textit{a}) the spanwise wavenumber $k_\textrm{max}$ at which $\|SG\|_\infty$ attains its maximum;
and in panel (\textit{b}) the efficiency $\alpha_\textrm{max}$ attained at this maximum.
We also plot these same quantities for the laminar velocity profiles (dashed lines) for comparison.
For both flows we see that $k_\textrm{max}$ remains almost constant as the friction Reynolds number is varied.
For Couette flow it lies in the range $1.18 \le k_\textrm{max} \le 1.19$; for Poiseuille flow it occurs at $k_\textrm{max}=1.62$ for all cases considered (including the laminar flow).
For both flows the efficiency $\alpha_\textrm{max}$ at this spanwise wavenumber is lower for the turbulent velocity profiles than it is for their laminar counterparts.
For Couette flow it reduces from $\alpha_\textrm{max}=0.98$ for the laminar profile to lie in the range $ 0.58 \le \alpha_\textrm{max} \le 0.63$ for the turbulent mean profiles.
For Poiseuille flow it reduces from $\alpha_\textrm{max}=0.32$ for the laminar profile to lie in the range $ 0.21 \le \alpha_\textrm{max} \le 0.28$ for the turbulent mean profiles.
Despite the greater reductions for Couette flow, $\alpha_\textrm{max}$ is nonetheless larger for Couette flow than it is for Poiseuille flow for all cases considered.
Indeed the maximum $\alpha_\textrm{max}$ attained for Poiseuille flow is still smaller than the minimum value attained for Couette flow (figure \ref{fig:effRe}\,\textit{b}).

\section{Singular value decomposition of $G$ and $S$ at $\omega=0$}
We have seen that the efficiency of the forcing process, as characterized by the quantity $\alpha$, is significantly higher for Couette flow than it is for Poiseuille flow for both laminar and turbulent mean velocity profiles.
We now seek to explain this observation by performing singular value decompositions of the Orr-Sommerfeld and Squire operators, $G$ and $S$.

As described in \S\,\ref{sec:norms}, the infinity norm \eqref{eq:infnorm} of a transfer function represents a maximum (or worst-case) gain.
This is attained at a particular forcing frequency and for a particular forcing direction, and therefore any analysis of $\|G\|_\infty$, $\|S\|_\infty$ and $\|F\|_\infty$ is complicated by the fact that each can occur at different temporal frequencies $\omega$.
For streamwise-constant fluctuations, however, the infinity norms $\|G\|_\infty$, $\|S\|_\infty$ and $\|F\|_\infty$ all occur at zero frequency, $\omega=0$.
We therefore need only consider each transfer function at $\omega=0$, i.e.~$G(\ii 0)$, $S(\ii 0)$ and $F(\ii 0)$.
A singular value decomposition of $G(\ii0)$ is then
\begin{equation}
  G(\ii0) = U \Sigma V^*,
\end{equation}
(and similarly for $S$) where $\Sigma = \textrm{diag}[\sigma_1 \, \cdots \, \sigma_n]$ contains the singular values with $\sigma_1 \ge \sigma_2 \ge \dots \ge \sigma_n$; $U = [u_1 \, \cdots \, u_n]$ contains the response singular vectors; and $V = [v_1 \, \cdots \, v_n]$ contains the forcing singular vectors.
Now writing the product of the two transfer functions $G$ and $S$ in terms of their singular value decompositions (using subscripts to distinguish between them):
\begin{equation} \label{eq:svd_product}
  F(\ii0) = S(\ii0) G(\ii0) = U_S \Sigma_S V_S^* U_G \Sigma_G V_G^*,
\end{equation}
from which we see that key roles will be played by $U_G$ (the singular response modes of $G$) and by $V_S$ (the singular forcing modes of $S$) since their product $V_S^* U_G$ appears at the centre of \eqref{eq:svd_product}.
The quantity $V_S^* U_G$ thus quantifies the nature of the interaction between the Orr--Sommerfeld and Squire operators, $G$ and $S$.

\subsection{Leading singular modes of $G$ and $S$}
Given the importance of $U_G$ and $V_S$ in determining the interaction between $G$ and $S$, we now plot their variation in the wall-normal direction.
We do so for Couette flow and Poiseuille flow and for their laminar and turbulent velocity profiles.
In all cases we set the spanwise wavenumber to $k_y = \pi/2$, which lies approximately in the range where the infinity norm of the overall system, $\|F\|_\infty$, achieves its largest value.
We plot only the first two singular modes of both operators because, as we will see, these are the most significant for understanding the results of \S\,\ref{sec:laminar} \& \ref{sec:turbulent}.

\begin{figure} \centering
  \includegraphics[width=1\textwidth]{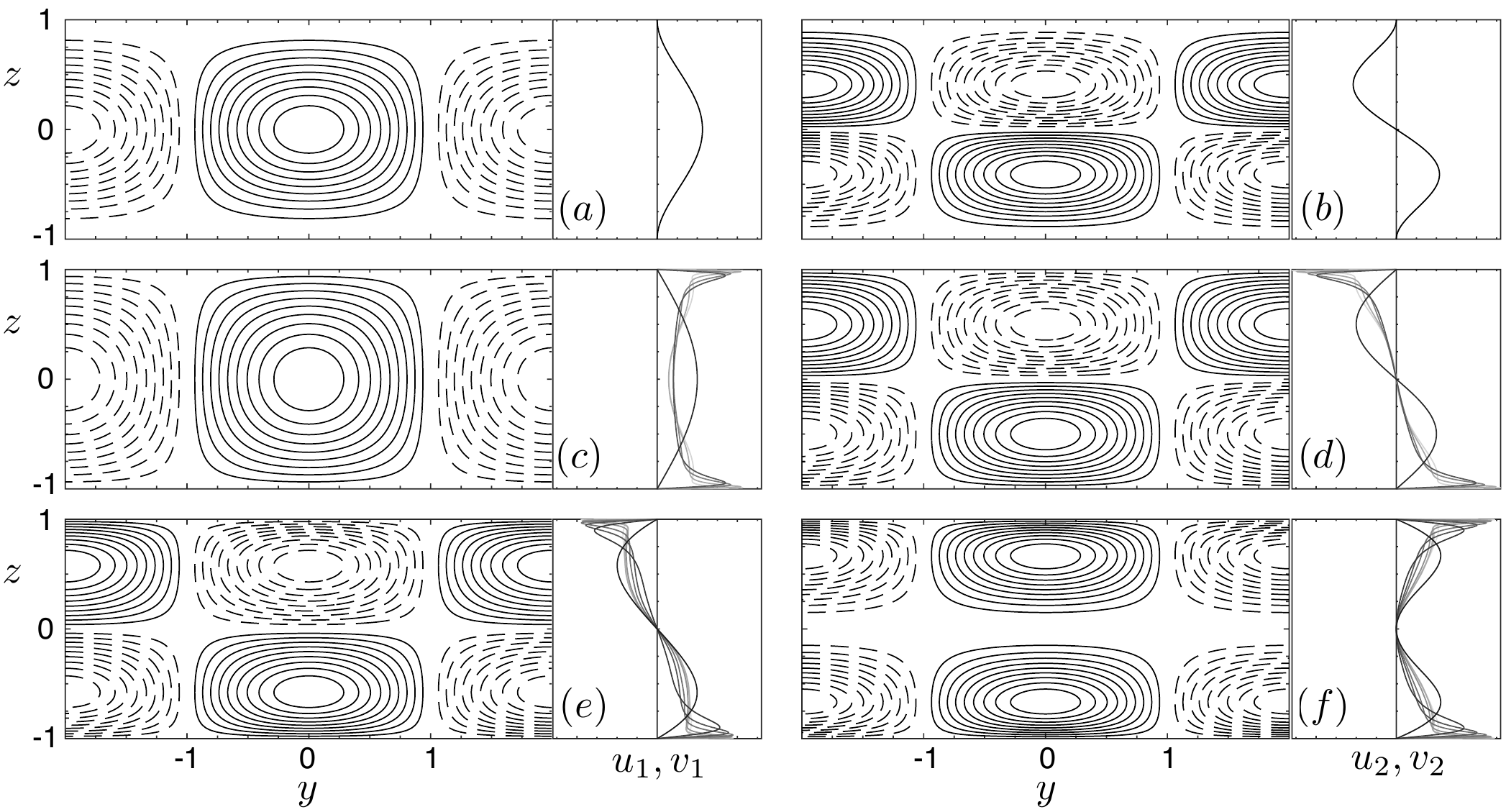}
  \caption{\label{fig:shapes}
    (\textit{a},\textit{b}) First two singular response modes ($u_1$,$u_2$) of $G$;
    and first two singular forcing modes ($v_1$ and $v_2$) of $S$ for Couette flow (\textit{c},\textit{d}) and Poiseuille flow (\textit{e},\textit{f}).
    In each panel the singular modes are shown in physical space on the left (laminar velocity profiles only); and in Fourier space on the right (laminar and turbulent velocity profiles; lighter lines correspond to larger friction Reynolds numbers).
    In each case the scaling is arbitrary.
  }
\end{figure}

The first two singular response modes ($u_1$ and $u_2$) of the Orr--Sommerfeld operator $G$ are shown in figure \ref{fig:shapes}\,(\textit{a},\textit{b}).
(These are plotted both in Fourier space and in physical space.)
Recall that, for streamwise-constant fluctuations, the Orr--Sommerfeld operator is independent of the mean velocity profile.
Therefore $u_1$ and $u_2$ remain the same across all of the velocity profiles that we consider (Couette and Poiseuille; laminar and turbulent).
The first response mode, $u_1$, spans the entire channel height and is symmetric about the channel centre (reaching its peak response there).
The second response mode, $u_2$, is anti-symmetric about the channel centre (where it is zero).

The first two singular forcing modes ($v_1$ and $v_2$) of the Squire operator $S$ for laminar Couette flow are shown in figure \ref{fig:shapes}\,(\textit{c},\textit{d}).
(Again, these are plotted both in Fourier space and in physical space.)
The first forcing mode $v_1$ is symmetric about the channel centre and is similar to $u_1$; and the second forcing mode $v_2$ is anti-symmetric about the channel centre and is similar to $u_2$.
This similarity between the leading forcing modes of $G$ and the leading response modes of $S$ can be quantified by looking at the inner products $v_1^*u_1$ and $v_2^*u_2$.
(This is equivalent to looking at the first two diagonal entries of $V_S^*U_G$.)
These are $v_1^*u_1 = 0.985$ and $v_2^*u_2 = 0.971$.
(The maximum possible value is 1 since $U_G$ and $V_S$ are each orthonormal.)
These values are also summarized in table \ref{tab:inner}.

The story is quite different for Poiseuille flow.
The first two singular forcing modes ($v_1$ and $v_2$) of the Squire operator $S$ for laminar Poiseuille flow are shown in figure \ref{fig:shapes}\,(\textit{e},\textit{f}).
The first forcing mode, $v_1$, is anti-symmetric about the channel centre; while the second forcing mode, $v_2$ is symmetric.
The anti-symmetry of $v_1$ means that its inner product with $u_1$ (which is symmetric) is zero, $v_1^*u_1=0$.
Thus the different symmetries of laminar Poiseuille flow mean that the leading forcing mode of the Squire operator is not excited by the leading response mode of the Orr-Sommerfeld operator. (The two modes are orthogonal.)
In a similar way, the symmetry of $v_2$ means that its inner product with $u_2$ (which is anti-symmetric) is zero, $v_2^*u_2=0$.
Thus for Poiseuille flow the important interactions are the second response mode with the first forcing mode ($v_1^* u_2 = 0.931$); and the first response mode with the second forcing mode ($v_2^*u_1=0.515$).
This limits the forcing efficiency of laminar Poiseuille flow and is a consequence of its anti-symmetrical shear profile $U'$.

Similar arguments apply when we replace the laminar velocity profiles of \S\,\ref{sec:laminar} with the turbulent mean velocity profiles of \S\,\ref{sec:turbulent}.
In particular the first forcing mode $v_1$ of the Squire operator is symmetric for turbulent Couette flow and anti-symmetric for turbulent Poiseuille flow; and
the second forcing mode $v_2$ is anti-symmetric for turbulent Couette flow and symmetric for turbulent Poiseuille flow.
These first two forcing modes are shown in figure \ref{fig:shapes}\,(\textit{c}--\textit{f}) for all Reynolds numbers considered alongside their laminar counterparts. (For brevity they are shown only in Fourier space.)
Thus the symmetries observed in the leading forcing modes of $S$ for laminar Couette flow and laminar Poiseuille flow are retained by their turbulent counterparts.
The most notable difference is that, for the turbulent mean velocity profiles, the leading forcing modes display peaks near the channel walls.
These peaks occur for both Couette flow and Poiseuille flow and they move closer to the wall as Reynolds number increases.
(This is not surprising given the crucial role of the mean wall-normal shear for the Squire operator---and that this shear becomes increasingly concentrated at the wall with increasing Reynolds number.)
This in turn causes a reduction in the inner products considered earlier.
For Couette flow the inner product $v_1^*u_1$ reduces from 0.985 (laminar) to 0.544 (Re$_\tau=986$); and the inner product $v_2^*u_2$ reduces from 0.971 (laminar) to 0.496 (Re$_\tau=986$).
For Poiseuille flow the inner product $v_2^*u_1$ reduces from 0.515 (laminar) to 0.385 (Re$_\tau=2003$); and the inner product $v_1^*u_2$ reduces from 0.931 (laminar) to 0.806 (Re$_\tau=2003$).
(These values are also summarized in table \ref{tab:inner}.)
This indicates (for both flows) a reduction in the projection of the leading response modes of $G$ onto the leading forcing modes of $S$ and helps to explain the reduced efficiency $\alpha$ for the turbulent mean velocity profiles observed in figures \ref{fig:conversionRe} \& \ref{fig:effRe}.

We finish this section by considering why laminar Couette flow displays almost perfect forcing efficiency, $\alpha \approx 1$?
This is explained by the fact that, for laminar Couette flow, the shear profile satisfies $U'(z)=1$ everywhere.
Thus the Squire operator becomes simply $S = (sI-\Delta)^{-1}$, the dynamics of which is similar to that of $(s\Delta-\Delta^2)^{-1}$, which appears in the Orr--Sommerfeld operator $G(s)$ (see \eqref{eq:G}).
It is then perhaps not surprising that the leading forcing modes of $S$ in figure \ref{fig:shapes}\,(\textit{a},\textit{b}) are so similar to the leading response modes of $G$ in figure \ref{fig:shapes}\,(\textit{c},\textit{d}).
More generally (not shown) the $i^\textrm{th}$ forcing mode of $S$ is similar to the $i^\textrm{th}$ response mode of $G$ and thus the inner product $v_i^* u_i$ is close to unity.

\begin{table}
  \centering
  \begin{tabular}{rrrrrrrrr}
    & \multicolumn{4}{c}{Couette} & \multicolumn{4}{c}{Poiseuille} \\
    \cmidrule(lr){2-5} \cmidrule(lr){6-9}
    & \multicolumn{2}{c}{laminar} & \multicolumn{2}{c}{$R_\tau=986$} & \multicolumn{2}{c}{laminar} & \multicolumn{2}{c}{$R_\tau=2003$} \\
    \cmidrule(lr){2-3}\cmidrule(lr){4-5}\cmidrule(lr){6-7}\cmidrule(lr){8-9}
    & \multicolumn{1}{c}{$v_1^*$} & \multicolumn{1}{c}{$v_2^*$} & \multicolumn{1}{c}{$v_1^*$} & \multicolumn{1}{c}{$v_2^*$} & \multicolumn{1}{c}{$v_1^*$} & \multicolumn{1}{c}{$v_2^*$} & \multicolumn{1}{c}{$v_1^*$} & \multicolumn{1}{c}{$v_2^*$} \\
    \midrule
    $u_1$     & 0.985 &     0     &     0.544 &     0     &         0 & 0.515     &         0 & 0.385 \\
    $u_2$     &     0 & 0.971     &         0 & 0.496     &     0.931 &     0     &     0.806 &     0 \\
    \bottomrule
  \end{tabular}
  \caption{Summary of the inner products $v_i^*u_j$ for $i,j=1,2$ (equivalent to looking at the first two diagonal entries of $V_S^*U_G$).}
  \label{tab:inner}
\end{table}

\section{Conclusions}
We have considered the linear amplification mechanisms leading to streamwise-constant large-scale structures in laminar and turbulent channel flows.
To do so the Orr--Sommerfeld and Squire operators have each been considered separately.
This is advantageous in three ways:
\textit{i}) it makes explicit the forcing of streamwise velocity fluctuations by wall-normal velocity fluctuations;
\textit{ii}) it allows one to define an efficiency of this forcing process; and
\textit{iii}) it exploits the fact that, for streamwise-constant fluctuations, the dynamics governing the wall-normal velocity are independent of the mean velocity profile (and therefore the mean shear).
The analysis helps to explain the prevalence of streamwise-constant structures with a spanwise spacing of approximately $4h$ (where $h$ is the channel half height) in both laminar and turbulent channel flows.
This spanwise spacing is encoded in the Orr-Sommerfeld operator $G(s)$ (figure \ref{fig:infnorm}) which, for streamwise-constant fluctuations, depends only on the channel geometry and boundary conditions.
The analysis also indicates that Couette flow is more efficient than Poiseuille flow in leveraging wall-normal velocity fluctuations to extract energy from the mean shear to produce streamwise velocity fluctuations.
This helps to explain the energetic large-scale roll modes observed in Couette flows over a wide range of Reynolds numbers.
As well as being of interest from a modelling point of view, the analysis could also have implications for control.
For example, one way to reduce the propensity for large-scale roll modes---as demonstrated by Poiseuille flow---is to choose boundary conditions such that the dynamics governing the wall-normal velocity and the dynamics governing the streamwise velocity have different symmetries.
While this might suggest possibilities that are ultimately impractical, at its best it could uncover possibilities that would otherwise be missed.

\section*{Acknowledgements}
I am grateful for the financial support of the Australian Research Council.

\appendix
\section{Two norm of $G$, $S$ and $F$ with spanwise wavenumber} \label{app:2norm}
\begin{figure} \centering
  \includegraphics[width=1\textwidth]{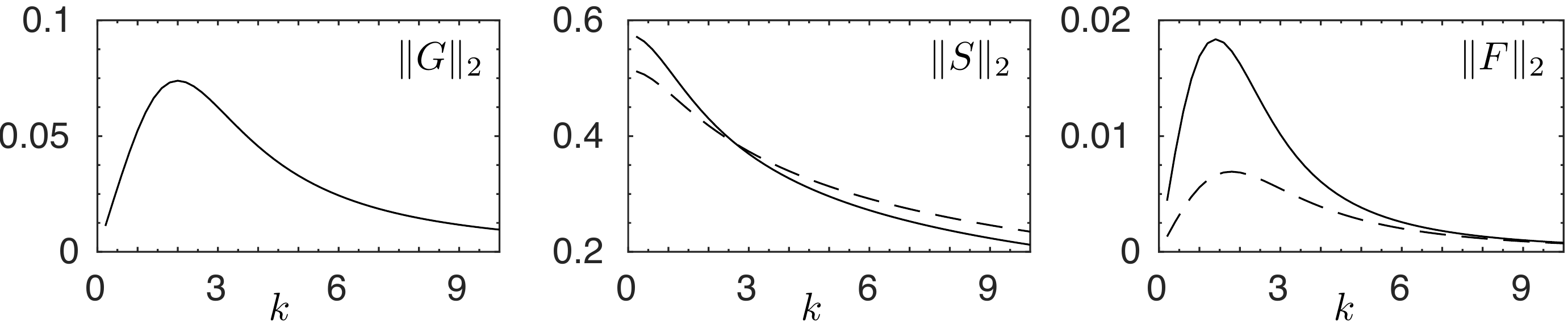}
  \caption{\label{fig:2norm}
    The two norm $\|\cdot\|_2$ for $G$ (left); $S$ (middle); and $F=SG$ (right).
    Results for $S$ and $F$ are shown for Couette flow (---) and for Poiseuille flow ($--$).
  }
\end{figure}

The two norm \eqref{eq:2norm} of $G$, $S$ and $F$ as a function of the spanwise wavenumber $k$ are plotted in figure \ref{fig:2norm}.
(The spanwise wavenumbers at which the peak in the two norm occurs are summarized in table \ref{tab:kmax} in \S\,\ref{sec:normk}.)

\bibliographystyle{jfm}
\bibliography{biblio}

\end{document}